\documentclass[aps,prd,eqsecnum,nofootinbib,showpacs,twocolumn]{revtex4-1}

\usepackage{amsmath,amssymb,bm}
\usepackage[dvips]{graphicx}
\usepackage{color}
\usepackage{textcomp}

\newcommand{\be}{\begin{equation}}
\newcommand{\ee}{\end{equation}}
\newcommand{\bea}{\begin{eqnarray}}
\newcommand{\eea}{\end{eqnarray}}

\newcommand{\eq}[1]{Eq.~(\ref{eq:#1})}
\newcommand{\sect}[1]{Sec.~\ref{sec:#1}}
\newcommand{\appen}[1]{Appendix~\ref{sec:#1}}

\newcommand{\del}{\partial}

\newcommand{\Tc}{T_c}

\newcommand{\zh}{z_h}
\newcommand{\zhc}{z_{h,c}}

\newcommand{\bra}{\langle}
\newcommand{\ket}{\rangle}
\newcommand{\calO}{{\cal O}}

\newcommand{\eg}{{\it e.g.}}
\newcommand{\ie}{{\it i.e.}}
\newcommand{\cf}{{\it c.f.}}

\newcommand{\bh}{black hole\ }
\newcommand{\EF}{Eddington-Finkelstein\ }

\newcommand{\HSC}{holographic superconductor\ }
\newcommand{\HSCs}{holographic superconductors\ }

\bmdefine{\bmq}{{\bm{q}}}
\bmdefine{\bmk}{{\bm{k}}}
\bmdefine{\bmx}{{\bm{x}}}
\bmdefine{\bmy}{{\bm{y}}}
\bmdefine{\bmr}{{\bm{r}}}
\bmdefine{\bmnabla}{{\bm{\nabla}}}
\bmdefine{\bmPsi}{{\bm{\Psi}}}

\newcommand{\calL}{{\cal L}}

\newcommand{\tiltau}{\tilde{\tau}}
\newcommand{\tilxi}{\tilde{\xi}}
\newcommand{\tiltkz}{\tilde{\tau}_\text{KZ}}
\newcommand{\tilxkz}{\tilde{\xi}_\text{KZ}}
\newcommand{\tilt}{\tilde{t}}
\newcommand{\tilv}{\tilde{v}}
\newcommand{\tilx}{\tilde{x}}
\newcommand{\tilbmx}{\tilde{\bmx}}
\newcommand{\tilz}{\tilde{z}}
\newcommand{\tilphi}{\tilde{\phi}}

\newcommand{\tilbmnabla}{\tilde{\bmnabla}}
\newcommand{\tilL}{\tilde{L}}

\newcommand{\vecx}{\vec{x}}

\newcommand{\tq}{\tau_Q}
\newcommand{\tkz}{\tau_\text{KZ}}
\newcommand{\xkz}{\xi_\text{KZ}}

\newcommand{\alphas}{a}
\newcommand{\betas}{b}
\newcommand{\gammas}{c}

\newcommand{\epsT}{\epsilon}
\newcommand{\epsmu}{\epsilon_{\mu}}

\newcommand{\nw}{\omega}
\newcommand{\nq}{q}

\newcommand{\tcL}{\Tilde{\calL}_\Psi}

\bmdefine{\bmA}{ \bm{A} }
\bmdefine{\bmD}{ \bm{D} }
\newcommand{\hcL}{ \calL_\Psi }

\newcommand{\barmu}{ \Bar{\mu} }

\begin{document}


\title{Kibble-Zurek scaling in holography}
\author{Makoto Natsuume}
\email{makoto.natsuume@kek.jp}
\altaffiliation[Also at]{
Department of Particle and Nuclear Physics, 
SOKENDAI (The Graduate University for Advanced Studies), 1-1 Oho, 
Tsukuba, Ibaraki, 305-0801, Japan;
 Department of Physics Engineering, Mie University, 
 Tsu, 514-8507, Japan.}
 \affiliation{KEK Theory Center, Institute of Particle and Nuclear Studies, 
High Energy Accelerator Research Organization,
Tsukuba, Ibaraki, 305-0801, Japan}
\author{Takashi Okamura}
\email{tokamura@kwansei.ac.jp}
\affiliation{Department of Physics, Kwansei Gakuin University,
Sanda, Hyogo, 669-1337, Japan}
\date{\today}
\begin{abstract}
The Kibble-Zurek (KZ) mechanism describes the generations of topological defects when a system undergoes a second-order phase transition via quenches. We study the holographic KZ scaling using holographic superconductors. The scaling can be understood analytically from a scaling analysis of the bulk action. The argument is reminiscent of the scaling analysis of the mean-field theory but is more subtle and is not entirely obvious. This is because the scaling is not the one of the original bulk theory but is an emergent one that appears only at the critical point. The analysis is also useful to determine the dynamic critical exponent $z$.
\end{abstract}
%

\maketitle

\section{Introduction}

The AdS/CFT duality \cite{Maldacena:1997re,Witten:1998qj,Witten:1998zw,Gubser:1998bc} has been useful to study ``real-world" which is rather difficult to analyze via conventional methods. The duality has been applied to QCD, condensed-matter physics, and nonequilibrium physics (see., \eg, Refs.~\cite{CasalderreySolana:2011us,Natsuume:2014sfa,Ammon:2015wua,Zaanen:2015oix} for textbooks). In this paper, we analytically study the holographic Kibble-Zurek (KZ) mechanism; The KZ mechanism describes the generations of topological defects when a system undergoes a second-order phase transition via quenches (cooling) \cite{Kibble:1976sj,Zurek:1985qw,Zurek:1996sj,Kibble:2007zz}. 

For a system with a second-order phase transition, the correlation length $\xi$ diverges, and as a result various physical quantities diverge. The divergences are parametrized by critical exponents, and those exponents define the universality class of the system. In the static case, thermodynamic quantities diverge but in the dynamic case, the relaxation time of the order parameter also diverges, which is known as the critical slowing down \cite{Hohenberg:1977ym,critical_text1,critical_text2}. The relaxation time $\tau$ behaves as $\tau \propto \xi^z$, where $z$ is a dynamic critical exponent. The study of dynamic critical phenomena in the AdS/CFT duality was pioneered by Refs.~\cite{Maeda:2008hn,Maeda:2009wv}.

When the system undergoes the phase transition through a quench, topological defects are generated spontaneously. As an example, for superconductors, the defects generated are vortices. By adding the quench, the symmetry is broken but spatially separated regions can select different states. The typical size of the correlated regions (domains) and the density of defects show power-law behaviors on the quench time scale $\tq$. This KZ mechanism is closely related to the critical slowing down. 

Recently, the holographic KZ mechanism has been studied numerically \cite{Sonner:2014tca,Chesler:2014gya}, and the KZ scaling has been confirmed. These works study \HSCs \cite{Gubser:2008px,Hartnoll:2008vx,Hartnoll:2008kx} which undergo second-order phase transitions. Typically, a \HSC is an Eintein-Maxwell-complex scalar system. For $T>\Tc$, the solution is a standard \bh without scalar, but for $T<\Tc$, the solution becomes unstable and is replaced by a solution with scalar ``hair."
According to Refs.~\cite{Sonner:2014tca,Chesler:2014gya}, after $t=\tkz$, where $\tkz$ is called the ``freeze-out time," one starts to have ``droplets," which are localized solutions of nonzero condensate. Droplets subsequently grow, and eventually droplets merge into $|\Psi| \simeq (\text{constant})$ solution leaving isolated regions with $\Psi=0$. They are vortices with winding number. Static holographic vortex solutions are obtained in Refs.~\cite{Albash:2009ix,Albash:2009iq,Montull:2009fe,Maeda:2009vf}. 

The holographic KZ scaling has been confirmed numerically. On the other hand, the KZ scaling can be understood from a scaling analysis of the mean-field theory such as the time-dependent Ginzburg-Landau theory (see, \eg, Ref.~\cite{Nikoghosyan:2013fqa}). Similarly, we would like to understand the holographic KZ scaling analytically. Our results are summarized as follows:
\begin{enumerate}
\item
The holographic KZ scaling can be understood analytically from a scaling analysis of the bulk action/equation of motion. The argument is reminiscent of the scaling analysis of the KZ mechanism.
\item
While the scaling analysis of the KZ mechanism, which is reviewed in \sect{KZM}, is straightforward, the holographic scaling analysis is more subtle and is not entirely obvious. This is because the scaling is not the one of the original bulk theory but is an emergent one which appears only at the critical point.
\end{enumerate}
Namely, the bulk equation of motion has the relativistic ``$z=1$" scaling, which comes from the underlying AdS geometry.
However, the operator $\hcL$ (\ref{eq:def-scalar-hcL_Psi}) of the complex scalar field,
which represents the time-independent homogeneous part of the equation of motion, has a zero eigenvalue at the critical point. In this case, the ``$z=2$" scaling,
which acts on the AdS radial coordinate trivially, is allowed. This ``$z=2$" scaling gives rise to the KZ scaling.

The emergent scaling is by no means surprising. Recall that the scaling in critical phenomena itself is an emergent one and is not transparent to see it from the underlying microscopic theory, \eg, from the Ising model. Similarly, a holographic analysis is a ``first-principle" computation in principle. The holographic scaling is emergent and is not transparent just like the scaling from the Ising model. 

As the result of the first-principle computation, a holographic analysis is usually not as easy as a mean-field analysis. A bulk system often has several fields and solving the equations of motion is not very easy. Thus, one often needs numerical analysis even to obtain mean-field results. 
However, a lesson of our analysis is that one does not have to solve the bulk system in order to obtain some qualitative information such as critical exponents and KZ scalings. The trick is that the eigenmode of the zero eigenvalue plays the role of the mean-field. 
There is a trade-off for analytic arguments however. We cannot really address the details of dynamical evolutions. They require numerical computations such as Refs.~\cite{Sonner:2014tca,Chesler:2014gya}.

The plan of this paper is as follows. In \sect{KZM}, we review the scaling argument to determine critical exponents and the KZ scaling. In \sect{HSC}, we set up our conventions for \HSCs especially in the \EF coordinates. The holographic counterpart of the scaling argument is given in \sect{HKZ}. We first use our scaling argument to rederive critical exponents $(\nu,z)=(1/2,2)$, which was shown in Ref.~\cite{Maeda:2009wv}. We then apply the holographic scaling argument to the KZ scaling. Finally, our work is related to various previous works \cite{Maeda:2009wv,Basu:2011ft,Basu:2012gg,Das:2014lda,Bhaseen:2012gg}, and we discuss those works in \sect{discussion}.

\section{KZ mechanism and scaling}\label{sec:KZM}

Consider the standard Ginzburg-Landau theory, or the $\phi^4$ mean-field universality class. The pseudofree energy $I[\phi]$ is given by 
\be
I[\phi] = \int d\bmx \,\left\{ \frac{1}{2}\,|\bmnabla \phi|^2 + \frac{\epsT}{2}\,|\phi|^2 + \frac{g}{4} |\phi|^4 \right\}~,
%
\ee 
where $\epsT := (T-\Tc)/\Tc$. Phenomenologically, the relaxation of a system is well-described by the time-dependent Ginzburg-Landau (TDGL) equation:
\be
\del_t\phi(t,\bmx) = - \int d\bmy\, \Gamma\big( |\bmx-\bmy| \big)\, \frac{\delta I[\phi]}{\delta \phi(t, \bmy)} + \zeta(t, \bmx)~,
%
\ee
where $\Gamma\big( |\bmx-\bmy| \big)$ is a transport coefficient, and $\zeta$ is a random Gaussian variable with
$ \bra \zeta(t,\bmx)\zeta(t',\bmx') \ket = 2T \Gamma(\bmx-\bmx')\delta(t-t') $.
The existence of $\zeta$ is crucial to produce fluctuations, but it does not play an important role below, so we henceforth ignore the term.

The details of dynamic universality classes partly depend on $\Gamma$. We shall focus on ``Model A" dynamic universality class \cite{Hohenberg:1977ym} because this is the dynamic universality class of \HSCs \cite{Maeda:2009wv}. This is the case where the order parameter is not a conserved charge. (On the other hand, the conserved charge case is the ``Model B" universality class.) In this case, $\Gamma$ is a constant,
and the TDGL equation becomes
\be
\del_t\phi = - \Gamma\left[ -\bmnabla^2 \phi + \epsT\,\phi + g \phi|\phi|^2 \right]~.
\label{eq:tdgl}
\ee

\subsection{Critical exponents from scaling}

We first determine critical exponents via a scaling argument.
Consider the scaling%
\footnote{The scaling of $\phi$ is chosen to be consistent with \eq{tdgl_scaling_choice}.}
\be
\tilt = \alphas t~, \quad \tilbmx = \betas \bmx~, \quad \tilphi = \phi/\betas~.
\label{eq:tdgl_scaling}
\ee
Under the scaling, the TDGL equation becomes
\be
\del_{\tilt} \tilphi = - \Gamma\left[ -\frac{\betas^2}{\alphas} \tilbmnabla^2 \tilphi + \frac{\epsT}{\alphas} \tilphi + \frac{\betas^2}{\alphas} g\tilphi|\tilphi|^2 \right]~.
\label{eq:tdgl_scaled}
\ee
The $(T-\Tc)$-dependence can be eliminated by choosing
\begin{subequations}
\label{eq:tdgl_scaling_choice}
\begin{align}
\alphas &= \epsT \propto (T-\Tc)~, \\
\betas &= \alphas^{1/2} \propto (T-\Tc)^{1/2}~.
%
\end{align}
\end{subequations}
Since the $(T-\Tc)$-dependence is eliminated, the system is away from the critical point in rescaled variables. Suppose the correlation length $\tilxi$ and the relaxation time $\tiltau$ in rescaled variables are $\tilxi \sim O(1)$ and $\tiltau \sim O(1)$. Then, in original variables,
\begin{subequations}
\begin{align}
\tau &= \alphas^{-1}\tiltau \propto (T-\Tc)^{-1}~,\\
\xi &= \betas^{-1}\tilxi \propto (T-\Tc)^{-1/2}~. 
%
\end{align}
\end{subequations}
The static exponent $\nu$ and the dynamic exponent $z$ are defined by
\begin{subequations}
\begin{align}
\tau &\propto \xi^z \propto |T-\Tc|^{-\nu z}~,\\
\xi &\propto  |T-\Tc|^{-\nu}~.
%
\end{align}
\end{subequations}
Thus, $(\nu, z)=(1/2,2)$ for Model A.

\subsection{KZ scaling}

We now consider the quench from high-temperature phase to low-temperature phase. One typically considers the linear ``quench protocol" 
\be
\epsT = -\frac{t}{\tq}~.
\label{eq:protocol_linear}
\ee
The quench is added for the initial temperature $T_i>\Tc$ to the final temperature $T_f<\Tc$ according to \eq{protocol_linear} so that the system crosses the critical point at $t=0$.

When the temperature change is slow enough compared with the relaxation time, the order parameter can adjust to the change, and the evolution is adiabatic. However, as we saw, the relaxation time of the order parameter diverges because of the critical slowing down. The evolution of the order parameter is slow, and the system cannot adjust to the change any more no matter how slow the quench is. The order parameter cannot adjust to the change globally and can adjust to the change only locally. As a result, defects form. This happens when $t <|\tkz|$, where $\tkz$ is called the ``freeze-out time." The size of the typical domain is called the ``freeze-out length" and is denoted as $\xkz$.

To determine $\tkz$ and $\xkz$, again consider the scaling \eqref{eq:tdgl_scaling}. In this case, the ``mass term" [the second term of the right-hand side of \eq{tdgl_scaled}] takes the form
\be
-\frac{ \tilt }{\alphas^2 \tq}\, \tilphi~,
%
\ee
so the $\tq$-dependence can be eliminated by choosing
\begin{subequations}
\begin{align}
\alphas &= \tq^{-1/2}~, \\
\betas &= \alphas^{1/2} = \tq^{-1/4}~.
%
\end{align}
\end{subequations}
In rescaled variables, the relaxation time $\tiltkz$ and the correlation length $\tilxkz$ do not depend on $\tq$. Then, in original variables,
\begin{subequations}
\begin{align}
\tkz &\propto \alphas^{-1} = \tq^{1/2}~, \\
\xkz &\propto \betas^{-1} = \tq^{1/4}~.
%
\end{align}
\end{subequations}
This agrees with the KZ prediction
\begin{subequations}
\begin{align}
\tkz &\propto \tq^{\nu z/(1+\nu z)}~, \\
\xkz &\propto \tq^{\nu/(1+\nu z)}~,
%
\end{align}
\end{subequations}
for Model A \cite{Zurek:1985qw}.

It is easy to generalize the scaling argument to the polynomial quench of the form
\be
\epsT = -\left|\frac{t}{\tq}\right|^n \text{sgn}(t)~.
%
\ee
In this case,
\begin{subequations}
\label{eq:tdgl_kz}
\begin{align}
\tkz &\propto \tq^{n/(n+1)}~, \\
\xkz &\propto \tq^{n/2(n+1)}~.
%
\end{align}
\end{subequations}

To summarize, scaling arguments eliminate the $\epsT$-dependence in the previous subsection and the $\tq$-dependence in this subsection, which determines critical exponents $(\nu,z)$ and the KZ exponents. We essentially repeat similar scaling arguments holographically in \sect{HKZ}.

\section{Holographic superconductor}\label{sec:HSC}

\subsection{Preliminaries}

We consider the $(p+2)$-dimensional $s$-wave holographic superconductors given by
\begin{align}
   S
  &= \int d^{p+2}x\, \sqrt{-g} \bigg[ R - 2 \Lambda 
\nonumber \\ & 
- \frac{1}{e^2} \left\{ 
  \frac{1}{4} F_{MN}^2 + \left\vert D \Psi \right\vert^2 + V\left( |\, \Psi\, |^2 \right) 
  \right\}
  \bigg]~,
\label{eq:full_action}
\end{align}
where 
\begin{align}
  & F_{MN} = 2\, \partial_{[M} A_{N]}~,
& & D_M := \nabla_M - i A_M~,
\label{eq:def-covariant_deri} \\
  & \Lambda
  = - \frac{p (p + 1)}{2\, L^2}~,
& & V= m^2 |\Psi|^2~.
\label{eq:def-potential_V}
\end{align}
We use capital Latin indices $M, N, \ldots$ for the $(p+2)$-dimensional bulk spacetime coordinates and use Greek indices $\mu, \nu, \ldots$ for the $(p+1)$-dimensional boundary coordinates. 

Below, we take the probe limit $e\gg1$, where the backreaction of the matter fields onto the geometry is ignored. Then, in the static case, the background metric is given by the Schwarzschild-AdS black hole:
\be
ds_{p+2}^2 
= \left(\frac{L}{u}\right)^2 \left(-f(u)dt^2+ d\vecx_p^2 + \frac{du^2}{f(u)} \right)~, 
\label{eq:SAdS}
\ee
where
\be
%
f(u) := 1-\left(\frac{u}{u_h}\right)^{p+1}~.
%
\ee
Here, the boundary coordinates are written as  $x^\mu = (t, \vecx) = (t, x^i)$, and $u=0$ at the AdS boundary. The Hawking temperature is given by $T=(p+1)/(4\pi u_h)$. We set $L=e=1$ below.

In the $A_u=0$ gauge, the asymptotic behavior of matter fields is given by 
\begin{subequations}
\label{eq:Psi-asympt}
\begin{align}
   \Psi(x, u)
  &\sim 
  \Psi^{(-)}(x)\, u^{\Delta_-}
    + \Psi^{(+)}(x)\, u^{\Delta_+}~,
\label{eq:Psi-asym_behavior} \\
   \Delta_\pm
  &:= \frac{p + 1}{2}
  \pm \sqrt{ \left( \frac{p + 1}{2} \right)^2 + m^2 }~,
\label{eq:def-Delta-Psi} \\
  A_\mu(x, u)
  & \sim A_\mu^{(0)}(x)
  + A_\mu^{(1)}(x)\, u^{p-1}~.
\label{eq:A-asym_behavior}
\end{align}
\end{subequations}
$\Psi^{(+)}$ represents the order parameter expectation value $\bra\calO\ket$, and 
$\Psi^{(-)}$ represents the external source for $\calO$.
Since we are interested in the spontaneous condensate, we set $\Psi^{(-)}=0$%
\footnote{For simplicity, we do not consider the ``alternative quantization," where the role of $\Psi^{(+)}$ and $\Psi^{(-)}$ is exchanged \cite{Klebanov:1999tb}.}.
Similarly, the fast falloff $ A_\mu^{(1)}$ represents the boundary current $\bra J^\mu \ket$, and the slow falloff $ A_\mu^{(0)}$ represents its source (the external chemical potential $\mu$ and vector potential).

On the horizon, we impose the regularity condition for a time-independent problem, and we impose the ``incoming-wave" boundary condition for a time-dependent problem. The boundary condition is discussed more in next subsection.

\subsection{Eddington-Finkelstein coordinates}\label{sec:EF}

It is convenient to introduce the tortoise coordinate $u_*$ and the ingoing Eddington-Finkelstein (EF) coordinate system $(v,z)$, where
\begin{align}
du_* &:= - \frac{du}{f}~, \\
v &:= t + u_*~, \quad z=u~.
%
\end{align}
The horizon is located at $u_*\to -\infty$. The metric becomes
\be
ds_{p+2}^2 = \frac{1}{z^2} \left(-f(z)dv^2 - 2dv dz + d\vecx_p^2 \right)~.
\label{eq:EF_SAdS}
\ee
The inverse metric is $g^{vz}=-z^2$ and $g^{zz}=z^2f$.

The Maxwell field components in the EF coordinate system are related to the ones in the Schwarzschild-like coordinate system in \eq{SAdS} as
\begin{subequations}
\label{eq:EF_Maxwell}
\begin{align}
A_v &= A_t~, \\
A_z &= A_u + \frac{A_t}{f}~,
%
\end{align}
\end{subequations}
since $A_t dt + A_u du = A_v dv +A_z dz$.

The Maxwell-scalar system admits a static solution%
\footnote{$A_z=0$ is our gauge choice, and it is different from choosing $A_u=0$ in the Schwarzschild-like coordinates from \eq{EF_Maxwell}.}
\begin{subequations}
\label{eq:EF_sol}
\begin{align}
\bmA_v &= \mu~\varphi(z/z_h)~, 
\quad \varphi(x) := 1 - x^{p-1}~, \\
\bmA_z &= \bmA_i = 0~, \\
\bm{\Psi} &=0~,
%
\end{align}
\end{subequations}
where boldface letters indicate background values.
However, at the critical point, the $\bmPsi=0$ solution becomes unstable and is replaced by a $\bmPsi\neq0$ solution. 
For $p=2$, $\Tc \approx 0.0587 \mu$. 

We approach the critical point from high temperature. When $\bmPsi=0$, the scalar perturbation decouples from the Maxwell perturbations, and it is enough to solve the scalar perturbation only.
The scalar action in the EF coordinate system is given by
\begin{subequations}
\begin{align}
   S_\Psi &= \int d^{p+2}x\,L
  ~,
\\
   L
  &= \frac{1}{z^p}\, \left[\, \big( \partial_v \Psi \big)^\dag\,
      \partial_z \Psi
    + \big( \partial_z \Psi \big)^\dag\, \partial_v \Psi
    - \delta^{ij}\, (\partial_i \Psi)^\dagger\, \partial_j \Psi\, \right] 
\nonumber \\
    &- \frac{1}{z^p} \left[ f \left| \left(\del_z+\frac{iA_v}{f}\right)\Psi \right|^2 
    +  \left( \frac{m^2}{z^2} 
    - \frac{A_v^2}{f} \right) |\Psi|^2 \right]
  ~.
\end{align}
\end{subequations}
For brevity, we write $\bmA_M$ as $A_M$. After integrating by parts,
\begin{subequations}
\label{eq:scalar-action-EF}
\begin{align}
   S_\Psi &= \int d^{p+2}x\,L
  ~,
\label{eq:scalar-action-SC-EF} \\
   L
  &= \frac{1}{z^p}\, \left[\, \big( \partial_v \Psi \big)^\dag\,
      \partial_z \Psi
    + \big( \partial_z \Psi \big)^\dag\, \partial_v \Psi
    - \delta^{ij}\, (\partial_i \Psi)^\dagger\, \partial_j \Psi\, \right] 
\nonumber \\ & 
  - \Psi^\dagger\, \hcL\, \Psi
\label{eq:scalar-L_Psi-I} \\
   \hcL
  &:= - \left( \partial_z \!+\! \frac{iA_v}{f} \right)
    \frac{f}{z^p} \left( \partial_z \!+\! \frac{iA_v}{f} \right) 
    + \frac{1}{z^p} \left( \frac{m^2}{z^2} 
    \!-\! \frac{A_v^2}{f} \right)~.
\label{eq:def-scalar-hcL_Psi}
\end{align}
\end{subequations}
The operator $\hcL$ plays the important role below. It represents the time-independent homogeneous part of the equation of motion. Thus, at the critical point, there must exist a nontrivial solution of $\hcL \Psi=0$ with $\Psi^{(-)}=0$%
\footnote{Actually, the $iA_v$-dependence in $\hcL$ can be gauged away, but we keep this form (until \sect{positivity}) so that the $A_v$-dependence in the action \eqref{eq:scalar-action-EF} is contained only in $\hcL$.}.

The EF coordinate system is convenient because the boundary condition at the horizon is simple. One often uses the Schwarschild-like coordinates and imposes the ``incoming-wave" boundary condition at the horizon. In the incoming EF-coordinates, the boundary condition reduces to the regularity condition.
In the near-horizon limit $z\to\zh$, the Lagrangian becomes
\be
L \propto -(\del_v\Psi)^\dag \del_*\Psi - (\del_*\Psi)^\dag \del_v\Psi + \Psi^\dag \del_*^2\Psi~,
%
\ee
and the equation of motion becomes
\be
\del_*(2\del_v+\del_*)\Psi \sim 0~.
%
\ee
There are two solutions.
The solution of $\del_*\Psi \sim 0$, namely $\Psi=\Psi(v)$ is the incoming-wave, and the other one is the outgoing-wave. 

In the EF-coordinates, the boundary condition becomes simple. This allows us to implement the holographic scaling analysis directly at the action level. 
In the Schwarzschild-like coordinates, this structure is manifest only after one imposes the ``incoming-wave" boundary condition explicitly. 

The action \eqref{eq:scalar-action-EF} allows the obvious ``$z=1$" scaling
\be
v \to \alphas v~, \quad
x^i \to \alphas x^i~, \quad
z \to \alphas z~.
\label{eq:scaling_obvious}
\ee
Under the scaling, the horizon radius and the chemical potential scale as 
\be
\zh \to \alphas \zh~, \quad \mu \to \mu/\alphas~,
%
\ee
but this does not change physics since the system is parametrized by the dimensionless parameter $\barmu := \zh\, \mu \propto \mu/T$. The scaling comes from the underlying AdS \bh geometry \eqref{eq:EF_SAdS}. Then, one would naively expect $z=1$, \ie, $\tau \sim \xi$, but at the critical point, one actually has $z=2$ as we see below. This is because of the emergent scaling at the critical point.
At the critical point, $\hcL \Psi=0$. This allows the scaling 
\be
v \to \alphas v~, \quad 
x^i \to \alphas^{1/2} x^i~, 
\quad z \to z~.
%
\ee

\section{Holographic scaling argument}\label{sec:HKZ}

Following the spirit of the TDGL scaling argument in \sect{KZM}, consider the scaling 
\be
\tilv = \alphas\, v~, \quad 
\tilx^i = \betas\, x^i~, \quad
\tilz = \gammas\, z~,
\label{eq:scaling}
\ee
in the scalar action \eq{scalar-action-EF}. 
\begin{itemize}
\item
In general, one would allow a $\Psi$-scaling like the TDGL analysis \eqref{eq:tdgl_scaling}, but our scalar action is quadratic in $\Psi$, so the $\Psi$-scaling is irrelevant in the discussion below. 
\item
The scaling $\gammas$ can be set to a desired value without loss of generality, and we choose convenient ones below.
\item
Only constant scalings are considered to keep a simple scaling property of the ``kinetic term" [the first line of \eq{scalar-L_Psi-I}].
\end{itemize}

\subsection{Assumptions}\label{sec:assumptions}

In the TDGL analysis, one can determine the $(T-T_c)$-dependence. A parallel discussion becomes possible by utilizing some generic properties of the eigenvalue problem
\be
\hcL \Psi = \lambda \Psi~.
%
\ee
The eigenvalue $\lambda(\barmu)$ depends on $\barmu := \zh\, \mu \propto \mu/T$. We impose the boundary conditions where $\Psi\sim z^{\Delta_+}$ asymptotically and is regular at the horizon. We impose the following assumptions on this eigenvalue problem:
\begin{enumerate}
\item
Under our boundary conditions, we assume that $\hcL$ has a discrete and positive spectrum above $\Tc$ (or below $\barmu_c$). 

\item
Denote the lowest eigenvalue of $\hcL$ as $ \lambda_0(\barmu)$. As one lowers $T$, a nontrivial source-free solution of $\hcL\Psi=0$ first appears at the critical point $\barmu = \barmu_c$, so $\lambda_0(\barmu_c) = 0$.

\item
The dynamics of $\Psi$ is governed by the $\lambda_0$-eigenmode near the critical point%
\footnote{Reference~\cite{Basu:2011ft} also emphasizes the importance of the zero eigenvalue mode when one discusses scaling properties in a quench problem.}.

\end{enumerate}
For the moment, we take these assumptions for granted, but we justify Assumptions 1 and 2 later. The (gauge-equivalent) operator is written as
\be
\hcL^g = \hcL(\barmu=0) - \frac{A_v^2}{z^p f}~.
%
\ee
We will show that $\hcL(\barmu=0)$ has positive-definite eigenvalues if $m^2$ satisfies the Breitenlohner-Freedman (BF) bound \cite{Breitenlohner:1982bm}. But the Maxwell field contribution is negative: this decreases the $\hcL$-eigenvalues when one increases $\barmu$, and $\hcL$ develops a zero eigenvalue at the critical point.


\subsection{Critical exponents from scaling}

In this subsection, we determine critical exponents $\nu$ and $z$. We first choose $\gammas$. To determine critical exponents, it is enough to consider the static background where the temperature ($\sim 1/\zh$) is constant, and we choose $\gammas=1/\zh$. Then, the coordinate $\tilz$ is dimensionless. The horizon is located at $\tilz_h=1$, and $f=1-\tilz^{p+1}$. 

The scaled action then becomes
\begin{subequations}
\label{eq:scalar-scaling-action-EF}
\begin{align}
   S_\Psi
  &= \frac{1}{ (z_h\, \betas)^p }\,\int d^{p+2}\tilx~\tilL
  ~,
\label{eq:scalar-scaling-action-SC-EF} \\
   \tilL
  &= \frac{1}{\tilz^p}\,
  \left[\, \big( \partial_{\tilv} \Psi \big)^\dag\, \partial_{\tilz} \Psi
  + \big( \partial_{\tilz} \Psi \big)^\dag\, \partial_{\tilv} \Psi 
\right. \nonumber \\ & \left. 
  -\frac{ z_h\, \betas^2 }{\alphas}\, \delta^{ij}\,
    \big( \Tilde{\partial}_{i} \Psi \big)^\dag\, \Tilde{\partial}_{j} \Psi
  \, \right]
  - \Psi^\dagger\,
    \bigg( \frac{\tcL}{z_h\, \alphas} \bigg)\, \Psi
  ~,
\label{eq:scalar-scaling-L_Psi-I} \\
   \tcL
  &= - \bigg( \partial_{\tilz}
    + i\, \barmu\, \frac{ \varphi(\tilz) }{f(\tilz)} \bigg)\,
    \frac{f}{\tilz^p}\, \bigg( \partial_{\tilz}
      + i\, \barmu\, \frac{ \varphi(\tilz) }{f(\tilz)} \bigg)
\nonumber \\ & 
  + \frac{1}{\tilz^p}\, \left\{ \frac{m^2}{\tilz^2}
    - \barmu^2\, \frac{ \varphi^2(\tilz) }{f(\tilz)} \right\}
  ~.
\label{eq:scalar-scaling-hcL_Psi-I}
\end{align}
\end{subequations}
The scaled action reduces to the original action if one chooses
\begin{subequations}
\begin{align}
\zh\, \alphas &= (\zh\, \betas)^2~, \\
\zh\, \alphas &=1~.
\label{eq:scaling_choice2}
\end{align}
\end{subequations}
Note that the second condition comes from the $\tcL$-term. These two conditions give $\alphas=\betas=1/\zh$, but it is just the obvious scaling \eqref{eq:scaling_obvious}.

However, $\tcL$ has a zero eigenvalue at the critical point, \ie, $\tcL\Psi=0$. Then, the story is different, and the anisotropic ``$z=2$" scaling $\alphas\propto\betas^2$ is allowed since one does not have to impose \eq{scaling_choice2}.

Instead, from our Assumption~3,
\be
  \Psi^\dag \frac{\tcL}{\zh\, \alphas} \Psi 
  \sim \frac{\lambda_0(\barmu)}{\zh\,\alphas} \Psi_0^\dag \Psi_0~,
%
\ee
near the critical point, where $\Psi_0$ is the $\lambda_0$-eigenmode. So, one can choose
\be
\zh\,\alphas = \lambda_0(\barmu)~.
%
\ee
In rescaled variables, the system is away from the critical point. Let the correlation length $\tilxi$ and the relaxation time $\tiltau$ in rescaled variables be $\tilxi \sim O(1)$ and $\tiltau \sim O(1)$. Then, in original variables,
\begin{subequations}
\begin{align}
\tau &= \alphas^{-1}\tiltau \propto \lambda_0^{-1}(\barmu)~, \\
\xi &= \betas^{-1}\tilxi \propto \alphas^{-1/2} \propto \lambda_0^{-1/2}(\barmu)~, \\
\to \tau &\propto \xi^2~.
%
\end{align}
\end{subequations}
The correlation length and the relaxation time diverge at the critical point as expected, and we obtain $z=2$. 

Note that the lowest eigenvalue $\lambda_0$ plays the role of $\epsT \propto T-\Tc$ in the TDGL scaling argument in \sect{KZM}. In both arguments, we eliminate the dependence of the deviation from the critical point by the scaling, and the scaling determines the critical exponents.

Denote the deviation from the critical point as 
\be
\epsmu := 1-\frac{\barmu}{\barmu_c}~.
%
\ee
The static exponent $\nu$ defined by $\xi \propto |\epsmu|^{-\nu}$ can be determined as follows. Expand $\tcL$ around $\barmu_c$ in the $\epsmu$-expansion.
The operator becomes $\hcL(\barmu_c)$ at the leading order, so it vanishes for $\Psi_0$.  At next order, the operator is proportional to $\epsmu$ (\cf, see the next subsection. We will carry out such an expansion.) Thus, $\lambda_0(\barmu) \propto |\epsmu|$, and
\be
%
\xi \propto \lambda_0^{-1/2} \propto  |\epsmu|^{-1/2}~,
%
\ee
namely $\nu=1/2$. Put differently, the $\barmu$-dependence in $\hcL$ is regular around $\barmu_c$, which allows the $\epsmu$-expansion. This is the essential reason why $\nu=1/2$.

\subsection{Holographic KZ scaling}

Now, consider the quench case. The system is parametrized by $\mu/T$, but a time-dependent chemical potential is physically meaningless, so we consider the time-dependent \bh temperature. 
Consider the quench protocol as
\begin{align}
z_h(v) & = \zhc \left\{ 1 + \text{sgn}(v)\, \left( \frac{|\, v\, |}{\tq} \right)^n \right\} \\
&= \zhc\{ 1- \epsmu(v) \}~,
\label{eq:quench-z_h}
\end{align}
where $\zhc$ is the horizon radius at the critical point. $\epsmu(v)$ represents the deviation from the critical point:
\be
\epsmu(v)
  := 1-\frac{z_h(v)}{\zhc}
  ~.
\label{eq:quench-bareps_mu}
\ee

One should actually consider a dynamical \bh and would change the \bh temperature $T$, or the horizon radius $\zh$ dynamically. Instead, we keep using the background \eqref{eq:EF_SAdS} with $\zh=\zh(v)$. The background is no longer an exact solution; the Einstein equation gets corrections of $O(\del_v\zh)$. In principle, one can construct a dynamical solution in the expansion of $\del_v\zh$. But our background remains a good approximation when the quench is slow enough. This is because
\be
\del_v\zh \propto \zhc/\tq
%
\ee
for our protocol. 
A similar remark also applies to the Maxwell solution \eqref{eq:EF_sol}. Alternatively, for the Maxwell solution, one can use an exact solution in the background \eqref{eq:EF_SAdS} with $\zh=\zh(v)$%
\footnote{Consider a time-dependent chemical potential $\mu(v)$ instead of constant $\mu$. This is not physical, and we just use a gauge degree of freedom. Then, $A_v=c \zh(v)(1-z/\zh(v))$ is a solution. In this case, $\mu(v) \propto \barmu_c\{1-\epsmu(v)\}$, and one has to take into account the $\epsmu$-correction in the following discussion. The scaled action \eqref{eq:scalar-scaling-quench-2-hcL_Psi-I} takes a slightly different form, but \eq{scalar-scaling-quench-2-hcL_Psi-II} remains valid.}.

In previous subsection, we chose $\gammas=1/\zh$. In the quench case, $\zh=\zh(v)$, and we should not choose $\gammas=1/\zh(v)$; this would change the form of the kinetic term \eqref{eq:scalar-L_Psi-I}. In the action \eqref{eq:scalar-action-EF}, the horizon radius $\zh(v)$ appears
only in $f$ and $\varphi$; it appears in the form of
\be
\frac{z}{z_h(v)}
  = \frac{z}{ \zhc }\, \frac{1}{ 1 - \epsmu(v) }
  = \frac{\tilz}{ \gammas\, \zhc }\, \frac{1}{ 1 - \epsmu(v) }~.
%
\ee
So, in this case, it is convenient to choose $\gammas=1/\zhc$. Again choose the scaling $\zhc\,\alphas =(\zhc\betas)^2$ and rewrite the scaled action in the $\epsmu$-expansion. Then, we eliminate the $\tq$-dependence (in $\epsmu$) by choosing $\alphas$ appropriately.

Then, the scaled action becomes
\begin{subequations}
\label{eq:scalar-scaling-quench-2-action-EF}
\begin{align}
   S_\Psi
  &= \frac{1}{ (\zhc\, \betas)^p }\,\int d^{p+2}\tilx~\tilL
  ~,
\label{eq:scalar-scaling-quench-2-action-SC-EF} \\ 
   \tilL
  &= \frac{1}{\tilz^p}\,
  \left[\, \big( \partial_{\tilv} \Psi \big)^\dag\, \partial_{\tilz} \Psi
  + \big( \partial_{\tilz} \Psi \big)^\dag\, \partial_{\tilv} \Psi 
\right. \nonumber \\ & \left. 
  - \delta^{ij}\, \big( \Tilde{\partial}_{i} \Psi \big)^\dag\,
      \Tilde{\partial}_{j} \Psi\, \right]
  - \Psi^\dagger\, \bigg( \frac{\tcL}{\zhc\, \alphas} \bigg)\,
    \Psi
  ~,
\label{eq:scalar-scaling-quench-2-L_Psi-I} 
\displaybreak \\
  \tcL
  &= - \bigg( \partial_{\tilz}
    + i\, \barmu_c\, \frac{ \varphi(z/z_h) }{f(z/z_h)} \bigg)\,
    \frac{f}{\tilz^p}\, \bigg( \partial_{\tilz}
    + i\, \barmu_c\, \frac{ \varphi(z/z_h) }{f(z/z_h)} \bigg)
\nonumber \\ &  
  + \frac{1}{\tilz^p}\, \left\{ \frac{m^2}{\tilz^2}
    - \barmu^2_c\, \frac{ \varphi^2(z/z_h) }{f(z/z_h)} \right\}
\label{eq:scalar-scaling-quench-2-hcL_Psi-I} \\
  &= - \bigg( \partial_{\tilz}
    + i\, \barmu_c\, \frac{ \varphi(\tilz) }{f(\tilz)} \bigg)\,
    \frac{f(\tilz)}{\tilz^p}\, \bigg( \partial_{\tilz}
    + i\, \barmu_c\, \frac{ \varphi(\tilz) }{f(\tilz)} \bigg)
\nonumber \\ &  
  + \frac{1}{\tilz^p}\, \left\{ \frac{m^2}{\tilz^2}
    - \barmu^2_c\, \frac{ \varphi^2(\tilz) }{f(\tilz)} \right\}
\nonumber \\
  &+ O\left( \epsmu(v) \right)
  ~,
\label{eq:scalar-scaling-quench-2-hcL_Psi-II}
\end{align}
\end{subequations}
where we expand $f$ and $\varphi$ in $\epsmu$ in \eq{scalar-scaling-quench-2-hcL_Psi-II}. The first and the second lines of \eq{scalar-scaling-quench-2-hcL_Psi-II} are $\tcL$ at the critical point, so it vanishes for $\Psi_0$.
Thus, near the critical point,
\begin{align}
\frac{ \tcL }{\zhc\, \alphas} \Psi
  \sim \frac{ \epsmu(v) }{\zhc\, \alphas}\, \Psi_0
  &= \frac{1}{\zhc\,\alphas} \left( \frac{|v|}{\tq} \right)^n \Psi_0 
\label{eq:op_near_critical} \\
  &= 
  \frac{\zhc^n}{ (\zhc\, \alphas)^{n+1} } \left( \frac{|\tilv|}{\tq} \right)^n \Psi_0
  ~.
\end{align}
The $\tq$-dependence can be eliminated by choosing $\alphas$ as
\begin{align}
  &\zhc\, \alphas
  = 
  \left( \frac{\zhc }{\tq} \right)^{ n/(n+1) }
  ~.
\label{eq:scalar-scaling-quench-alpha}
\end{align}
In rescaled variables, the relaxation time $\tiltkz$ and the correlation length $\tilxkz$ do not depend on $\tq$. Then, in original variables,
\begin{subequations}
\label{eq:scalar-scaling-quench-KZ-all}
\begin{align}
  \tkz
  &\propto \alphas^{-1}
  \propto \tq^{ n/(n+1) }
  ~,
\label{eq:scalar-scaling-quench-KZ-tau} \\
  \xkz
  &\propto \betas^{-1} \propto \alphas^{-1/2}
  \propto \tq^{ n/2(n+1) }
  ~.
\label{eq:scalar-scaling-quench-KZ-xi}
\end{align}
\end{subequations}
This agrees with the TDGL analysis \eqref{eq:tdgl_kz}.

To be precise, in the above discussion, $\Psi_0$ is the $\lambda_0$-eigenmode at $v=0$. We impose the regularity boundary condition at the dynamical horizon $z=\zh(v)$. Thus, the boundary condition is imposed at a different position for a different $v$. See \appen{BC} for more details.

\subsection{Checking assumptions}\label{sec:positivity}

Let us go back to Assumptions~1 and 2 in \sect{assumptions} and justify them. We consider the eigenvalue problem
\begin{subequations}
\begin{align}
  & {\cal L}_\Psi(\barmu) \Psi = \lambda(\barmu) \Psi~,
\\
  & {\cal L}_\Psi
  := - \left( \del_z + \frac{iA_v}{f} \right)
    \frac{f}{z^p} \left( \del_z + \frac{i\, A_v}{f} \right)
  + \frac{ V_\Psi }{z^p}~,
\\
  & V_\Psi := \frac{m^2}{z^2} - \frac{A_v^2}{f}~.
\end{align}
\end{subequations}
What one can show is that 
\begin{enumerate}
\item $\lambda_{N} := \lambda(\barmu = 0) > 0$ if $m^2$ satisfies the BF bound.
\item The lowest eigenvalue $\lambda_0(\barmu \ne 0)$
is less than $\lambda_{N0} := \lambda_0(\barmu=0)$
because of the $A_v^2$-term in $V_\Psi$.
\end{enumerate}
First, the operator $\hcL$ is hermitian with respect to $\Psi$ which satisfy our boundary conditions, so the eigenvalues are real. The spectrum is also discrete as a two-boundary-value problem. 

It is easier to reduce to an equivalent eigenvalue problem. First, the $iA_v$-dependence in $\hcL$ can be gauged away by $\Psi = U \Psi^g$ where $U := \exp( - i \int dz\, A_v/f )$:
\begin{align}
\hcL^g \Psi^g &:= (U^\dag \hcL U) \Psi^g = \lambda \Psi^g~, \\
\hcL^g &= - \del_z\frac{f}{z^p}\del_z + \frac{ V_\Psi }{z^p}~.
%
\end{align}
Further define a new variable $\psi$ as $\Psi^g = G(z) \psi$. The function $G$ is chosen below. Then, the problem reduces to
\be
\calL_\psi \psi := (G \calL_\Psi^g G) \psi = \lambda G^2 \psi~.
\label{eq:eigen_full}
\ee
Showing $\lambda_N > 0$ amounts to showing the integral
\be
  \lambda = \frac{ \int dz\, \psi^\dag \calL_\psi \psi}
    { \int dz\, G^2 |\psi|^2 }
%
\ee
is positive-definite for $\barmu = 0$.
In the variable $\Psi$, the potential $V_\Psi$ has the $m^2<0$ term, so the positivity is not explicit even for $\barmu=0$. But in the new variable $\psi$, the positivity can be seen explicitly.
For simplicity, $\psi$ is normalized as $\int dz\, G^2 |\psi|^2=1.$

By integrating by parts,
\begin{align}
   \int dz\, \psi^\dag \calL_\psi \psi
  &= \int \frac{dz}{z^p} \bigg[ fG^2 |\del_z\psi|^2 
  + V_\psi G^2 |\psi|^2 \bigg]
\nonumber \\ &  
  + (\text{surface term})~.
\label{eq:RQ}
%
\end{align}
%
The kinetic term is positive-definite. 
The surface term makes no contribution under our boundary conditions. The new potential term is given by
\be
  V_\psi := \frac{m^2}{z^2}
  - \frac{z^p}{G} \left( \frac{fG'}{z^p} \right)'
  - \frac{A_v^2}{f}
  ~.
%
\ee
Choose a polynomial form of $G(z)=z^{\alpha/2}$ with constant $\alpha$. Then,
\begin{align}
   V_\psi
  &= \frac{1}{z^2} \bigg\{ - \frac{(\alpha - p - 1)^2}{4}
  + \left(\frac{p+1}{2}\right)^2 + m^2
\nonumber \\
  &\qquad
  + \frac{\alpha^2}{4}\, \left( \frac{z}{\zh} \right)^{p+1} \bigg\}
  - \frac{A_v^2}{f}
  ~.
%
\end{align}
By choosing $\alpha= p+1$, $V_\psi(\barmu=0)$ is positive-definite if $m^2 > -(p+1)^2/4$, \ie, if $m^2$ satisfies the BF bound. Thus, the integral \eqref{eq:RQ} is positive-definite for $\barmu = 0$,
and $\lambda_N > 0$.

Now, consider the $\barmu\neq0$ problem.  
Since $V_\psi < V_\psi(\barmu=0)$, $\lambda_0 < \lambda_{N0}$.
For convenience, we repeat the argument in Ref.~\cite{Basu:2010fa}.
Let the lowest eigenfunctions of $\calL_\psi (\barmu)$ and
$\calL_\psi(\barmu=0)$ be $\psi_0$ and $\psi_{N0}$. Then,
\begin{align}
\lambda_{N0} &= \int dz\, \psi_{N0}^\dag \calL_\psi(\barmu=0)\, \psi_{N0} 
\nonumber \\
&>  \int dz\, \psi_{N0}^\dag \calL_\psi(\barmu)\, \psi_{N0} 
\nonumber \\
&\geq  \int dz\, \psi_0^\dag \calL_\psi(\barmu)\, \psi_0 = \lambda_0(\barmu)~,
%
\end{align}
where the variational argument is used in the last line.

Although $\lambda_{N0} > 0$, $\lambda_0(\barmu)$ may not be
because of the Maxwell field contribution.
It is then natural to expect that $\lambda_0(\barmu)$ remains positive
for a small $\barmu$, but $\lambda_0(\barmu)$ tends to decrease
as one increases $\barmu$, and $\lambda_0=0$ for a large enough $\barmu_c$,
which is the critical point.


\section{Relation to previous works}\label{sec:discussion}

Our work is related to various previous works, and it is worthwhile to mention them and to compare with them briefly.

Reference~\cite{Maeda:2009wv} obtained critical exponents for \HSCs analytically. The analysis is carried out in the Schwarzschild-like coordinates and in the momentum space. But, in this paper, we are interested in defect formations, or solutions which are inhomogeneous in boundary spatial directions, so it is more appropriate to work in the real space. 
In any case, the main points of the paper are (i) the existence of a nontrivial solution with $\Psi^{(-)}=0$ at the critical point, or $\calL_\Psi\Psi=0$, and (ii) after imposing the boundary condition at the horizon, the scalar equation of motion has $O(\omega)$ and $O(q^2)$ terms.

More explicitly, the paper considers the solution of the form $\Psi = \Psi_{\nw, \nq}(u) e^{- i \omega t + i q x}$. In order to implement the ``incoming-wave" boundary condition, write the solution as
\begin{align}
  & \Psi_{\nw, \nq}(u)
  =: (1 - u)^{ - i \omega/(4 \pi T) } \varphi_{\nw, \nq}(u)~.
\label{eq:def-varphi}
\end{align}
Then, the $\varphi$ equation has $O(\omega)$ and $O(q^2)$ terms. This allows us to expand $\varphi$ as
$
\varphi_{\omega,q} = \varphi_{0} + \nw \varphi_{(1,0)}
          + \nq^2 \varphi_{(0,1)} + \cdots.
$
The main objet computed in the paper is the ``order parameter response function" $\chi_{\omega,q}$ given by
\begin{subequations}
\begin{align}
   \chi_{\nw, \nq}
  & \propto \frac{ \Psi^{(+)}_{\nw, \nq} }{ \Psi^{(-)}_{\nw, \nq} }
\\
  & \propto
    \frac{ \varphi_{0}^{(+)} + \nw \varphi_{(1,0)}^{(+)}
          + \nq^2 \varphi_{(0,1)}^{(+)} + \cdots }
         { \varphi_{0}^{(-)} + \nw \varphi_{(1,0)}^{(-)}
          + \nq^2 \varphi_{(0,1)}^{(-)} + \cdots }
 \\
  &\sim \frac{ \varphi_{0}^{(+)} }{ \varphi_{(0,1)}^{(-)} }~
    \frac{1}{- \frac{i}{\Gamma} \nw + \nq^2 + \frac{1}{\xi^2} }~,
\label{eq:response_fn}
\end{align}
\end{subequations}
where%
\footnote{According to an explicit numerical computation, $\Gamma$ is complex, so the order parameter is not purely diffusive \cite{Maeda:2009wv}.}
\be
\frac{1}{\Gamma} := i \frac{\varphi_{(1,0)}^{(-)}}{\varphi_{(0,1)}^{(-)}}~,
\quad
\frac{1}{\xi^2} := \frac{\varphi_{0}^{(-)}}{\varphi_{(0,1)}^{(-)}}~.
\frac{}{}
%
\ee
From \eq{response_fn}, one obtains $(\gamma,\nu,\eta,z)=(1,1/2,0,2),$ where the point (i), namely $\varphi_0^{(-)} |_{\Tc} = 0$, $\varphi_0^{(+)} |_{\Tc} \neq 0$ is essential. 

The quench in this paper is cooling, the standard quench discussed in the context of the KZ mechanism. Such a quench is called a ``thermal quench." This quench is added by the time-dependent \bh temperature $T(t)$. But a different type of quench is discussed in the literature. They typically consider the time-dependent source $\Psi^{(-)}(t)$ for the order parameter. Such a quench is called a ``source quench."

For example, Das and his collaborators consider source quenches at $T=0$ \cite{Basu:2011ft,Basu:2012gg,Das:2014lda}. Their analysis does not address defect formations for two reasons. First, the boundary theory is spatially homogeneous. Second, a source quench drives a spontaneously symmetry breaking system to an explicit symmetry breaking system. We consider the second-order phase transition and defect formations, so it is not appropriate to consider a source quench. Although their problem is not a defect formation problem, it is fine as a quench problem in a broad sense, and they use a similar scaling argument as ours.


A source quench for \HSCs is analyzed numerically in Ref.~\cite{Bhaseen:2012gg}. Their quench protocol has a Gaussian profile, so it drives the system to an explicit symmetry breaking one only in a limited time. They start from the ordered phase, add the source quench, and follow the time-evolution of $\bra\calO(t)\ket$. The system ends up with the disordered phase if the source quench is strong enough.

\begin{acknowledgments}

We would like to thank Takeshi Morita for useful discussions. 
This research was supported in part by a Grant-in-Aid for Scientific Research (23540326 and 17K05427) from the Ministry of Education, Culture, Sports, Science and Technology, Japan. 

\end{acknowledgments}

\newpage
\appendix 

\section{Time-dependent boundary condition and its effect}\label{sec:BC}

We impose the boundary condition at the dynamical horizon $\tilz=\zh(v)/\zhc=1-\epsmu(v)$. The boundary condition thus has the $O(\epsmu)$-dependence. Consequently, $\tcL$-eigenfunctions differ for a different $v$. 
The $\lambda_0$-eigenmodes get only $O(\epsmu)$-corrections, however. Namely, $\Psi_{0}(v) \sim \Psi_{0}(v=0)+O(\epsmu)$, and \eq{op_near_critical} remains valid. 

Alternatively, one can modify the argument so that the boundary condition has no explicit $\epsmu$-dependence. Then, the $\epsmu$-dependence is contained entirely in the operator $\tcL$, and the eigenfunctions remain the same for all $v$.
To do so, introduce a new variable $\sigma = z/z_h(v)$, and rewrite $\tcL$ in terms of $\sigma$. Then, the boundary condition is always imposed at $\sigma=1$. 

Introducing $\sigma$ is not a coordinate transformation. A $v$-dependent coordinate transformation would spoil the form of the kinetic term \eqref{eq:scalar-L_Psi-I}. Writing in terms of $\sigma$ is just a convenient way to shift the effect of the time-dependent boundary condition to the operator $\tcL$. As the $\tcL$-eigenvalue problem, one can regard $v$ just as an external parameter. 

Then, the effect of the time-dependent boundary condition is incorporated in the scaled action $\tcL$ as
%
%
%
\begin{subequations}
\label{eq:scalar-scaling-quench-2_v2-action-EF}
\begin{align}
  & \tcL
  = \left( \frac{ z_{h,c} }{z_h} \right)^{p+2}\,
  \Bigg[
  \nonumber \\ 
  &- \left\{ \del_{\sigma}
    + i \left( \frac{\barmu_c z_h}{ z_{h,c} } \right)
      \frac{ \varphi(\sigma) }{f(\sigma)} \right\}
    \frac{f(\sigma)}{\sigma^p}
  \left\{ \del_{\sigma}
    + i \left( \frac{\barmu_c z_h}{ z_{h,c} } \right)
      \frac{ \varphi(\sigma) }{f(\sigma)} \right\}
\nonumber \\
  &
  + \frac{1}{\sigma^p}\, \left\{ \frac{m^2}{\sigma^2} 
    - \left( \frac{\barmu_c z_h}{ z_{h,c} } \right)^{2}
      \frac{ \varphi^2(\sigma) }{f(\sigma)} \right\}
  \, \Bigg]
\label{eq:scalar-scaling-quench-2_v2-hcL_Psi-I} 
\\
  &= \left( \frac{ z_{h,c} }{z_h} \right)^{p+2}
  \Bigg[-\left\{ \del_{\sigma}
    + i \barmu_c \frac{ \varphi(\sigma) }{f(\sigma)} \right\}
    \frac{f(\sigma)}{\sigma^p}
  \left\{ \del_{\sigma}
    + i \barmu_c \frac{ \varphi(\sigma) }{f(\sigma)} \right\}
  \nonumber \\ 
  &+ \frac{1}{\sigma^p}\, \left\{ \frac{m^2}{\sigma^2}
    - \barmu^2_c \frac{ \varphi^2(\sigma) }{f(\sigma)} \right\}
  \, \Bigg]
\nonumber \\
  &
+ O\left( \epsmu(v) \right)
 ~,
\label{eq:scalar-scaling-quench-2_v2-hcL_Psi-II}
\end{align}
\end{subequations}
where we expand $\tcL$ in $\epsmu$ as in \eq{scalar-scaling-quench-2-hcL_Psi-II}. The prefactor $(\zhc/\zh)^{p+2}$ can also be expanded in $\epsmu(v)$. It is not necessary however since the first term vanishes for $\Psi_0$ at the critical point. 

The rest of the discussion remains the same as the text, and we get \eq{op_near_critical}.

\end{document}